\documentclass[preprint,prd,showpacs,preprintnumbers,amsmath,amssymb,floatfix]{revtex4-2}

\usepackage{amsmath}
\usepackage{graphicx}% Include figure files
\usepackage{dcolumn}% Align table columns on decimal point
\usepackage{bm}% bold math
\usepackage{ulem} %% for strike-through
\usepackage[usenames]{color}
\usepackage{epstopdf}
\usepackage{float}
\usepackage{multirow}

\usepackage{array}
\usepackage{booktabs}   % 用于美化表格线条
\usepackage{subcaption} % 更现代、功能更强的子图宏包，用于替换 subfigure
\usepackage{tabularx}

\newcommand{\nc}{\newcommand}       % new command
\nc{\vc}[1] {\mbox{\boldmath $#1$}} % boldmath(vector)
\nc{\del}       {\partial}              % bra state
\nc{\bra}       {\langle}               % bra state
\nc{\ket}       {\rangle}               % ket state
\nc{\bras}[1]   {\langle #1|}           % bra state
\nc{\kets}[1]   {|#1\rangle}            % ket state
\nc{\mapleft}[1]{           % something under arrow
	\smash{\mathop{\,          %
			\hbox to 1.5cm{\rightarrowfill}\, }\limits_{#1}}}
\nc{\nn}      {\\\nonumber} \nc{\vs}      {\vspace{-0.275cm}}
\nc{\fra}    {\frac{1}{2}}
\nc{\mb}        {\mathbf}

\usepackage[colorlinks,linkcolor=blue,anchorcolor=teal,citecolor=red]{hyperref}

\begin{document}
	
	\preprint{}
	\title{Role of the $\delta$ Meson in Softening the Symmetry Energy within the DDRHF Model} 
	
	\author{Qirui Li}
	\affiliation{School of Physics, Nankai University, Tianjin 300071,  China}
	
	\author{Jinniu Hu}
	\email{hujinniu@nankai.edu.cn}
	\affiliation{School of Physics, Nankai University, Tianjin 300071,  China}
	\affiliation{Shenzhen Research Institute of Nankai University, Shenzhen 518083, China}
	
	\author{Ying Zhang}
	\email{yzhangjcnp@tju.edu.cn}
	\affiliation{Department of Physics, Faculty of Science, Tianjin University, Tianjin 300072, China}
	
	\author{Hong Shen}
	\affiliation{School of Physics, Nankai University, Tianjin 300071,  China}
	
	\date{\today}
		
		\begin{abstract}
			We investigate the effects of the isovector-scalar $\delta$ meson on the density dependence of the symmetry energy within the density-dependent relativistic Hartree--Fock (DDRHF) framework. As a baseline, we generate $1006$ accepted DDRHF parametrizations including the $\sigma$, $\omega$, $\rho$, and $\pi$ mesons by imposing empirical constraints on the saturation properties of nuclear matter. The resulting symmetry-energy slope parameters are confined to relatively large values, $L\simeq65$--$110~\mathrm{MeV}$. Two representative parametrizations, denoted RHF-NK1 and RHF-NK2, are randomly selected from this ensemble. Starting from these two parametrizations, we introduce the $\delta$ meson and readjust the meson--nucleon couplings under the same saturation-property constraints. The numerical optimization shows that small values of $L$ are obtained most efficiently when the $\delta$ coupling is taken to be constant. In this case, $L$ is reduced from approximately $73$ to $32~\mathrm{MeV}$, while the binding energy per nucleon, saturation density, symmetry energy, and incompressibility coefficient remain nearly unchanged. A channel-by-channel decomposition shows that the softening is not caused by the direct $\delta$-meson contribution alone, but by a redistribution among the $\delta$, $\rho$, and $\pi$ mesons together with the isoscalar Fock contributions. The resulting neutron-star mass--radius relations shift toward smaller radii, indicating that the $\delta$ meson provides an efficient additional degree of freedom for controlling the isovector properties of DDRHF models.
		\end{abstract}
		
			\maketitle
		
		
	%\tableofcontents
	
	
	
	\section{Introduction}
	\label{introduction}
	Covariant density functional theory (CDFT) provides a self-consistent relativistic framework for describing finite nuclei and nuclear matter in terms of Lorentz-covariant energy density functionals. Its basic concepts, applications to finite nuclei and exotic systems, and modern extensions beyond the static mean-field approximation have been reviewed extensively in Refs.~\cite{ring1996relativistic,vretenar2005relativistic,meng2006relativistic,niksic2011relativistic}. Within this framework, relativistic Hartree--Fock (RHF) theory explicitly includes Fock terms in the relativistic many-body description of nuclei. Early RHF models developed by Brockmann \textit{et al.} and Bouyssy \textit{et al.} were applied to both nuclear matter and finite nuclei from the late 1970s to the mid-1980s~\cite{brockmann1978relativistic,horowitz1983properties,bouyssy1987relativistic}. To improve the description of bulk nuclear properties, nonlinear meson self-interactions were subsequently introduced~\cite{bernardos1993relativistic}. Microscopic RHF calculations with effective density-dependent meson exchange were then explored by Fritz \textit{et al.}, who mapped Dirac--Brueckner correlations based on realistic forces onto finite-nucleus RHF calculations and demonstrated the importance of Fock terms~\cite{fritz1993dirac}. The density-dependent RHF approach was further formulated by Shi \textit{et al.}, who parametrized Dirac--Brueckner--Hartree--Fock self-energies through density-dependent meson--nucleon couplings within the RHF framework~\cite{shi1995relativistic}. Long \textit{et al.} later developed this framework in a more systematic DDRHF formulation and constructed the effective interactions PKO1, PKO2, and PKO3~\cite{long2006density}. They also constructed the PKA1 interaction, which improves the description of shell structure and tensor-force effects~\cite{long2007shell}.
	\par
	{The DDRHF framework is adopted here because it explicitly retains exchange terms and finite-range meson-exchange channels. This is particularly relevant from the viewpoint of modern nuclear forces and chiral effective field theory (EFT), where pion exchange, constrained by chiral symmetry, provides the long-range spin-isospin and tensor components of the nuclear interaction. In Hartree-CDFT, density-dependent relativistic mean-field (DDRMF), and point-coupling models, the pion does not contribute at the Hartree level because of parity conservation in the nuclear ground state, and its effects are therefore not treated as an explicit exchange channel. By contrast, DDRHF includes the pion through Fock terms and also allows the rho-tensor and isoscalar exchange contributions to be separated. This makes DDRHF a complementary covariant framework for identifying how the delta meson modifies the balance among different isovector and exchange channels.}
	\par
	The DDRHF framework has also been applied to dense matter and neutron stars. Sun \textit{et al.} showed that Fock terms in PKO-type equations of state affect the symmetry energy, proton fraction, direct-Urca threshold, and mass--radius relation~\cite{sun2008neutron}. Subsequent DDRHF studies considered the hadron--quark phase transition, hyperon softening, tensor effects, $\Delta(1232)$ isobars, and higher-order symmetry-energy terms~\cite{sun2008hadronquark,sun2009neutronCPC,long2012hyperon,jiang2015tensor,zhu2016delta,liu2018fourthorder}. These works show that exchange terms can strongly influence the isospin composition and stiffness of dense matter. The density dependence of the nuclear symmetry energy therefore provides an essential link between microscopic isovector interactions and macroscopic neutron-star observables.
	\par
	The symmetry energy is constrained by finite nuclei, heavy-ion collisions, and neutron-star observations~\cite{li2008recent,baldo2016nuclear,li2021progress}. Around saturation density, neutron skins and electric dipole polarizabilities constrain the slope parameter $L$~\cite{adhikari2021accurate,adhikari2022precision,reed2021implications,reinhard2022combined,tamii2011complete,birkhan2017electric}. Heavy-ion observables, including nucleon emission ratios, light-cluster yields, isospin transport, collective flows, and meson production, provide complementary information from sub-saturation to supra-saturation densities~\cite{famiano2006neutron,li2006double,tsang2009constraints,chen2003light,zhang2005probing,tsang2004isospin,chen2005determination,mallik2022constraining,ciampi2022first,ciampi2025model,li2002probing,ferini2006isospin,russotto2016results,estee2021probing}. These constraints are valuable but remain affected by uncertainties in transport dynamics and reaction modeling.
	\par
	Neutron stars provide a natural laboratory for probing the high-density EOS. The maximum mass, tidal deformability, and mass--radius relation impose complementary constraints~\cite{lattimer2001neutron,fortin2016neutron}. GW170817 supplied the first gravitational-wave constraint on neutron-star tidal deformability and radii~\cite{abbott2017gw170817,abbott2018gw170817eos}. NICER pulse-profile analyses have measured the mass--radius distributions of PSR J0030+0451 and PSR J0740+6620, with updated results reported recently~\cite{riley2019psrj0030,miller2019psrj0030,riley2021j0740,miller2021j0740,vinciguerra2024updated,salmi2024radius}. The mass--radius data used below also include PSR J0437--4715, PSR J0614--3329, XTE J1814--338, and the compact object in HESS J1731--347~\cite{reardon2024neutron,choudhury2024nicer,mauviard2025nicer,kini2024constraining,doroshenko2022strangely}. These observations generally favor radii around $10$--$13~\mathrm{km}$ for low- and intermediate-mass neutron stars, although individual inferences remain model dependent.
	\par
	Recent data-driven analyses using astronomical observations with nuclear-model or causality-informed priors also favor relatively small canonical radii~\cite{zhou2023nonparametric,dong2025equation}. Zhou \textit{et al.} obtained $R_{1.4}=12.31^{+0.29}_{-0.31}~\mathrm{km}$ and $12.30^{+0.35}_{-0.37}~\mathrm{km}$ in a nonparametric deep-neural-network reconstruction, while Dong \textit{et al.} found $R_{1.4}\simeq11.85~\mathrm{km}$ with speed-of-sound constraints~\cite{zhou2023nonparametric,dong2025equation}. In contrast, representative PKO-type DDRHF interactions predict $R_{1.4}=13.5$--$14.5~\mathrm{km}$~\cite{sun2008neutron}. Since $R_{1.4}$ correlates strongly with the symmetry-energy slope parameter $L$~\cite{hu2020effects}, this comparison suggests that the conventional DDRHF isovector sector is too stiff.
	\par
	{Microscopic nuclear-matter calculations based on chiral EFT and ab initio many-body methods provide an important reference for the density dependence of the symmetry energy. In these approaches, pion exchange, constrained by chiral symmetry, gives the long-range spin-isospin and tensor components of the nuclear force. A Bayesian analysis of chiral-EFT neutron-matter calculations with correlated uncertainties gives $E_{\rm sym}=31.7\pm1.1$ MeV and $L=59.8\pm4.1$ MeV at saturation density~\cite{Drischler2020PRL}, while recent analyses constrained by different chiral-EFT calculations typically give $L$ values in the range of about $40$--$70$ MeV~\cite{Drischler2021ARNPS,Lim2024PRC}. This provides a useful benchmark for covariant energy-density functionals.  However, the present DDRHF parametrizations, such as PKO, and PKA, were confined to a larger range, $L\simeq75$--$100$ MeV. This motivates us to introduce an additional isovector degree of freedom to reduce $L$.}
		
		{The $\delta$ meson has been widely discussed in RMF and density-dependent RMF models. Its direct Hartree contribution to the symmetry energy is negative, although the refit of the $\rho$ channel may modify the high-density behavior~\cite{Kubis1997PLB}. Density-dependent and Bayesian RMF studies further show that the $\delta$ channel can enlarge the accessible parameter space of the symmetry-energy slope and curvature and affect neutron-star radii, tidal deformabilities, and proton fractions~\cite{RocaMaza2011PRC,Wang2014PRC,Teodoro2025PRC}. The present work examines whether a similar softening of the symmetry energy can be achieved in DDRHF and whether the mechanism is caused by the direct $\delta$ contribution or by the rearrangement of Fock channels.}
	
	\par
	Motivated by these considerations, we first examine whether a substantially smaller $L$ can be obtained in the no-$\delta$ DDRHF framework by readjusting the conventional $\sigma$, $\omega$, $\rho$, and $\pi$ channels under nuclear-matter saturation constraints. The isovector-scalar $\delta$ meson is then introduced as an additional degree of freedom. This channel has been widely discussed in RMF and density-dependent RMF models, including DD-ME$\delta$ and recent scalar-mixing extensions, for its impact on the neutron-proton effective-mass splitting, the density dependence of the symmetry energy, the proton fraction of $\beta$-stable matter, and neutron-star observables~\cite{kubis1997nuclear,kubis1998neutron,liu2002asymmetric,gaitanos2004lorentz,wang2014neutron,rocamaza2011relativistic,thakur2022effects,santos2025impact,li2022effects,kumar2023implications}. Guided by these studies, we analyze whether the $\delta$ meson can open a softer isovector sector in the DDRHF framework.
	\par
	The remainder of this paper is organized as follows. In Sec.~2, we briefly introduce the theoretical framework and outline the calculation of neutron-star properties. The results are presented and discussed in Sec.~3, followed by a summary in Sec.~4.
	\section{Theoretical framework}
	\label{sec:framework}
	The density-dependent relativistic Hartree--Fock (DDRHF) model is a meson-exchange framework in which relativistic effects are treated explicitly. In the present work, the isoscalar channel is mediated by the scalar meson ($\sigma$) and the vector meson ($\omega$), which account predominantly for the attractive and repulsive components of the nuclear interaction, respectively. The isovector sector includes the vector meson ($\rho$), the pseudoscalar meson ($\pi$), and the scalar meson ($\delta$). These channels govern the isospin dependence of the nuclear interaction, while the effects of the $\delta$-meson channel constitute the main focus of this work. Since only infinite nuclear matter is considered, the electromagnetic interaction is neglected. The corresponding Lagrangian density can be written as
	\begin{equation}
		\begin{aligned}	
			\mathcal{L}
			=
			&\bar{\psi}
			\left[
			i\gamma^\mu \partial_\mu
			- M
			- g_\sigma \sigma
			- g_\omega \gamma^\mu \omega_\mu
			- g_\rho \gamma^\mu \vec{\tau}\cdot \vec{\rho}_\mu
			+\frac{f_\rho}{2M}\sigma_{\mu\nu}\partial^{\nu}\vec{\rho}^{\mu}\right.\\
			&\left.\cdot\vec{\tau}- \frac{f_\pi}{m_\pi}\gamma_5\gamma^\mu \partial_\mu \vec{\pi}\cdot \vec{\tau}
			-g_{\delta}\vec{\delta}\cdot \vec{\tau}\right]\psi+ \frac{1}{2}\partial^\mu \sigma \partial_\mu \sigma
			- \frac{1}{2}m_\sigma^2 \sigma^2
			- \\
			&
			\frac{1}{4}\Omega^{\mu\nu}\Omega_{\mu\nu}
			+ \frac{1}{2}m_\omega^2 \omega_\mu \omega^\mu- \frac{1}{4}\vec{R}_{\mu\nu}\cdot \vec{R}^{\mu\nu}+ \frac{1}{2}m_\rho^2 \vec{\rho}^{\,\mu}\cdot \vec{\rho}_\mu+ \frac{1}{2}\partial_\mu \vec{\pi}\\
			&
			\cdot \partial^\mu \vec{\pi}
			- \frac{1}{2}m_\pi^2 \vec{\pi}\cdot \vec{\pi}+ \frac{1}{2}\partial_\mu \vec{\delta}\cdot \partial^\mu \vec{\delta}
			- \frac{1}{2}m_\delta^2 \vec{\delta}\cdot \vec{\delta}.
		\end{aligned}
	\end{equation}
	Here, \(\Omega^{\mu\nu}=\partial^\mu\omega^\nu-\partial^\nu\omega^\mu\) and \(\vec{R}^{\mu\nu}=\partial^\mu\vec{\rho}^{\,\nu}-\partial^\nu\vec{\rho}^{\,\mu}\) are the vector-meson field tensors.
	
	The meson--nucleon coupling constants are assumed to depend on the baryon density. For the $\sigma$- and $\omega$-meson couplings, the density dependence is parametrized as
	\begin{eqnarray}
		g_\phi(\rho_b) &=& g_\phi(\rho_0)f_\phi(\xi),
	\end{eqnarray}
	where $\phi=\sigma,\omega$. $\rho_0$ denotes the saturation density of nuclear matter, and $\xi=\rho_b/\rho_0$. The function $f_{\phi}(\xi)$ is defined as
	\begin{eqnarray}
		f_\phi(\xi) &=& a_\phi\frac{1+b_\phi(\xi+d_\phi)^2}{1+c_\phi(\xi+d_\phi)^2}.
	\end{eqnarray}
	
	For the $\rho$-, $\pi$-, and $\delta$-meson couplings, the density dependence is taken to be
	\begin{eqnarray}
		g_\rho(\rho_b) &=& g_\rho(0)e^{-a_\rho \xi}, \nonumber \\
		f_\rho(\rho_b) &=& f_\rho(0)e^{-a_{\rho f} \xi}, \nonumber \\
		f_\pi(\rho_b) &=& f_\pi(0)e^{-a_\pi \xi}, \nonumber \\
		g_\delta(\rho_b) &=& g_\delta(0)e^{-a_\delta \xi},
	\end{eqnarray}
	respectively. The parameter $a_\delta$ controls the density dependence of the $\delta$ coupling and is examined numerically below. The symbol $\xi$ is used here to distinguish the density variable in the coupling functions from the expansion variable used below for nuclear-matter coefficients.

	In the static approximation, the meson fields are eliminated through their equations of motion, leading to an effective nucleonic Hamiltonian with Hartree and Fock contributions from the individual exchange channels. Applying the variational principle gives the self-consistent RHF equation
	\begin{equation}
		(-i\gamma^\mu\partial_\mu+M+\Sigma)\psi(x)=0,
	\end{equation}
	where \(\Sigma\) is the nucleon self-energy. The no-sea approximation is adopted, and the rearrangement contribution induced by the density-dependent couplings is included in \(\Sigma\).

	Solving the RHF equation self-consistently, the ground-state energy density of nuclear matter is decomposed as
	\begin{equation}
		\varepsilon
		=
		\varepsilon_k + \sum_i \left(\varepsilon_i^D + \varepsilon_i^E\right).
	\end{equation}
	Here, $i=\sigma,\omega,\rho,\pi,\delta$ labels the different meson-exchange channels, with all $\rho$-mediated terms grouped into the $\rho$ channel. The quantity $\varepsilon_{k}$ denotes the kinetic contribution, while $\varepsilon_{i}^{D}$ and $\varepsilon_{i}^{E}$ represent the direct (Hartree) and exchange (Fock) contributions associated with the (i)-th interaction channel, respectively.
	
	The bulk properties are obtained from the energy per particle through the expansion
	\begin{eqnarray}
		e(\rho_b,\alpha) &=& e_{0}(\rho_b)+E_{\mathrm{sym}}(\rho_b)\alpha^2+\mathcal{O}(\alpha^4), \nonumber \\
		e_{0}(\rho_b) &=& E_{\mathrm{sat}}+\frac{1}{2}K_{\mathrm{sat}}x^2+\frac{1}{6}Q_{\mathrm{sat}}x^3+\cdots, \nonumber \\
		E_{\mathrm{sym}}(\rho_b) &=& E_{\mathrm{sym}}+Lx+\frac{1}{2}K_{\mathrm{sym}}x^2+\frac{1}{6}Q_{\mathrm{sym}}x^3+\cdots.
		\label{eq:bulk_expansion}
	\end{eqnarray}
	Here, $e=\varepsilon/\rho_b$ is the energy per baryon, $\rho_b=\rho_n+\rho_p$ is the baryon density, and
	$\alpha=\frac{\rho_n-\rho_p}{\rho_b}$
	denotes the isospin asymmetry. The dimensionless variable
	$x=\frac{\rho_b-\rho_0}{3\rho_0}$
	measures the deviation from the saturation density $\rho_0$. The coefficients $E_{\mathrm{sat}}$, $K_{\mathrm{sat}}$, and $Q_{\mathrm{sat}}$ characterize the saturation energy, incompressibility, and skewness of symmetric nuclear matter, respectively, while $E_{\mathrm{sym}}$, $L$, $K_{\mathrm{sym}}$, and $Q_{\mathrm{sym}}$ denote the symmetry energy and its slope, curvature, and skewness parameters at saturation density. Because of the definition of $x$, the conventional slope parameter is obtained directly from Eq.~\eqref{eq:bulk_expansion} as
	\begin{equation}
		L=\left.\frac{\partial E_{\mathrm{sym}}(\rho_b)}{\partial x}\right|_{x=0}.
		\label{eq:Ldef}
	\end{equation}
	For the channel decomposition discussed below, we use the corresponding density-dependent slope function
	\begin{equation}
		L(\rho_b)=3\rho_b\frac{\partial E_{\mathrm{sym}}(\rho_b)}{\partial \rho_b}
		=\frac{\rho_b}{\rho_0}\frac{\partial E_{\mathrm{sym}}(\rho_b)}{\partial x},
		\label{eq:Lrho}
	\end{equation}
	which reduces to Eq.~\eqref{eq:Ldef} at $\rho_b=\rho_0$.
	
	For neutron-star matter, the core is assumed to consist of neutrons, protons, electrons, and muons in $\beta$ equilibrium under charge neutrality. The leptons are treated as free Fermi gases. At a given baryon density, these conditions determine the particle fractions and, together with the baryonic contribution, the total energy density and pressure of the core. Because the meson--nucleon couplings depend on density, the rearrangement contribution is included consistently in the single-particle chemical potentials and in the pressure. The pressure is evaluated from the thermodynamic relation $P=\rho_b^2\partial(e)/\partial\rho_b$, equivalently $P=\sum_i\mu_i\rho_i-\varepsilon$, using the same chemical potentials that enter the $\beta$-equilibrium condition $\mu_n-\mu_p=\mu_e=\mu_\mu$. For the crust, we adopt the BPS and SLY4 equations of state~\cite{Haensel:1993zw,baym1971ground,douchin2001unified}. The mass--radius relation of a static, spherically symmetric neutron star is then obtained by solving the Tolman--Oppenheimer--Volkoff equations~\cite{oppenheimer1939massive,tolman1939static},
	\begin{eqnarray}
		\frac{dP}{dr}
		&=&
		-\frac{\left[P(r)+\varepsilon(r)\right]\left[M(r)+4\pi r^{3}P(r)\right]}
		{r\left[r-2M(r)\right]}, \nonumber \\
		\frac{dM}{dr}
		&=&
		4\pi r^{2}\varepsilon(r).
	\end{eqnarray}
	where $P(r)$, $\varepsilon(r)$, and $M(r)$ denote the pressure, energy density, and enclosed gravitational mass at radius $r$, respectively. Natural units $c=G=1$ are used. The above expressions present only the essential theoretical framework. A detailed derivation can be found in Refs.~\cite{long2006density,long2007shell,sun2008neutron}.

	\section{Results and discussion}
	\label{sec:results}
	We first quantify the no-$\delta$ baseline to separate the effect of adding the isovector-scalar channel from a simple refit of the conventional DDRHF interaction. The no-$\delta$ parametrizations contain the $\sigma$, $\omega$, $\rho$, and $\pi$ channels and are constrained by the empirical ranges
	\begin{eqnarray}\label{sat}
		30 \leq L \leq 120~\mathrm{MeV}, \nonumber \\
		30 \leq E_{\mathrm{sym}} \leq 34~\mathrm{MeV}, \nonumber \\
		220 \leq K \leq 280~\mathrm{MeV}, \nonumber \\
		-16.5 \leq E/A \leq -15.5~\mathrm{MeV}, \nonumber \\
		0.15 \leq \rho_{0} \leq 0.165~\mathrm{fm}^{-3}.
	\end{eqnarray}
	Here, $L$, $E_{\mathrm{sym}}$, $K$, $E/A$, and $\rho_0$ denote the symmetry-energy slope parameter, the symmetry energy at saturation density, the incompressibility coefficient, the binding energy per nucleon, and the saturation density, respectively.
	\par
\begin{figure}[t]
		\centering
		\includegraphics[width=0.6\columnwidth]{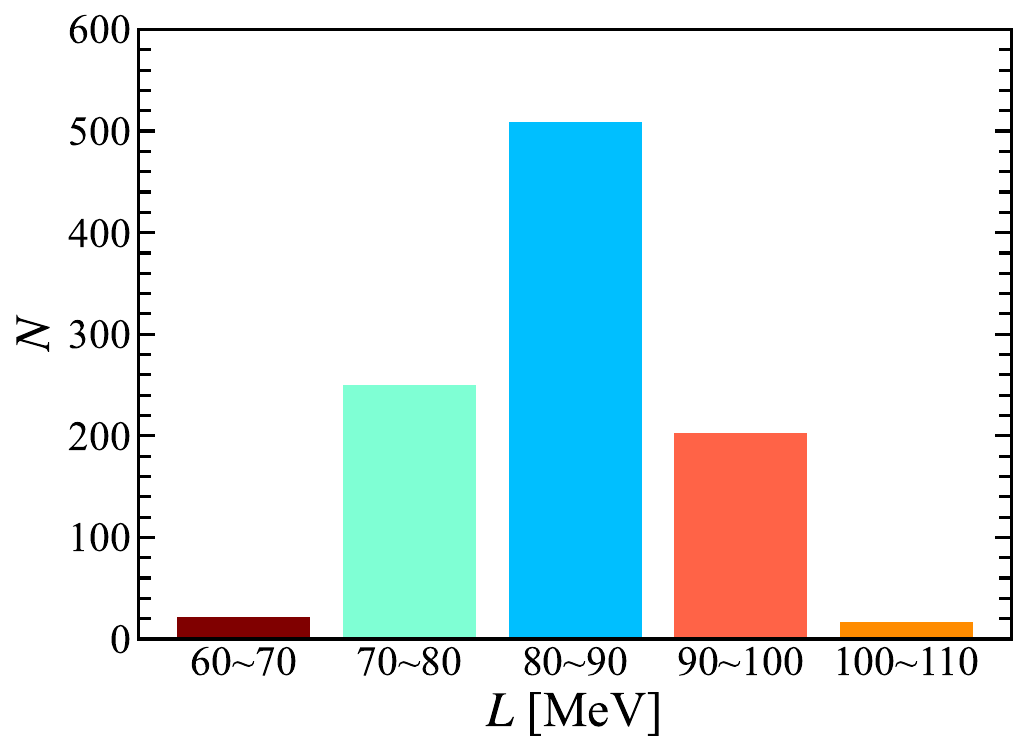}
		\caption{Distribution of the symmetry-energy slope parameter $L$ obtained from the $1006$ accepted no-$\delta$ DDRHF parameter sets constrained by the empirical saturation properties of nuclear matter.}
		\label{fig:L_distribution}
	\end{figure}

	Under these constraints, we obtain $1006$ accepted no-$\delta$ DDRHF parameter sets, {whose  saturation density, binding energy per nucleon, incompressibility coefficient, symmetry energy, and symmetry-energy slope all fall within the empirical intervals listed in Eq. \eqref{sat}}. Their symmetry-energy slope parameters, shown in Fig.~\ref{fig:L_distribution}, cover $L\simeq65$--$110~\mathrm{MeV}$, but no set enters the low-$L$ region, here referring to $L\lesssim 60$ MeV. This ensemble result indicates that the difficulty of obtaining a small $L$ is tied to the restricted isovector structure of the conventional DDRHF model rather than to a particular parametrization. The strong positive contributions from the $\sigma$ and $\omega$ mesons play an important role in this behaviour~\cite{zhao2015kin}. We therefore introduce the $\delta$ meson and refit the couplings under the same constraints.
	\par
	
	\par
	From this no-$\delta$ ensemble, two parametrizations are randomly selected and denoted RHF-NK1 and RHF-NK2. They are used as representative baselines for examining the effect of the $\delta$ meson. After the $\delta$ channel is incorporated and the couplings are readjusted, both parametrizations acquire much smaller values of $L$ while maintaining the empirical saturation properties.
	\par		Table~\ref{tab:coupling_parameters} shows the changes in the meson--nucleon couplings after the $\delta$ meson is introduced. The low-$L$ refit enhances the $\delta$ coupling, reduces the $\omega$- and $\pi$-meson couplings, and readjusts the $\rho$ channel. The value of {$g_\rho(0)$} increases, whereas the density-dependence parameter $a_\rho$ decreases. These changes indicate that the reduction of $L$ is not caused by the direct $\delta$ contribution alone, but by a redistribution among several meson-exchange channels.
	
\begin{figure}[t]
		\centering
		\includegraphics[width=0.6\columnwidth]{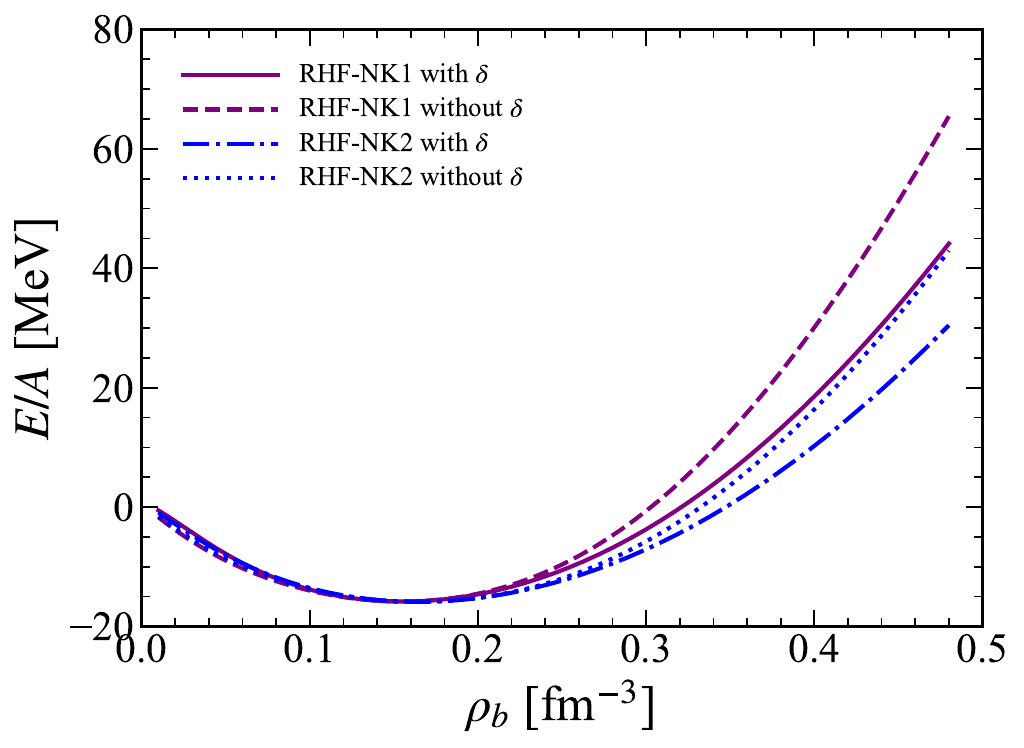}
		\caption{Energy per particle of symmetric nuclear matter for RHF-NK1 and RHF-NK2 before and after the inclusion of the $\delta$-meson channel.}
		\label{fig:eos_snm}
	\end{figure}

		{To further illustrate how the refit affects the isoscalar sector, we show in Fig.~\ref{fig:eos_snm} the energy per particle of symmetric nuclear matter , $E/A$, for RHF-NK1 and RHF-NK2 before and after the inclusion of the $\delta$ meson. Around the saturation density, the $E/A$  with and without the $\delta$ meson almost overlap. At higher densities, however, visible differences appear. In particular, the inclusion of the $\delta$ meson lowers the high-density energy per particle for RHF-NK1 and also modifies the RHF-NK2 curve moderately. Although the $\delta$ meson does not contribute directly to symmetric nuclear matter at the Hartree level, its introduction changes the optimized balance among the meson--nucleon couplings. The resulting readjustment of the isoscalar and exchange channels softens the high-density behavior while keeping the saturation region essentially unchanged.}
	
	\begin{table}[!htbp]
		\centering
		\caption{
			Meson--nucleon coupling parameters before and after
			the inclusion of the $\delta$-meson channel. A dash in the
			``with $\delta$'' column means that the corresponding
			parameter is kept unchanged from the no-$\delta$ value.}
		\label{tab:coupling_parameters}
		\footnotesize
		
		\setlength{\tabcolsep}{3pt}
		
		\renewcommand{\arraystretch}{1.05}
		\begin{tabular}{lcccc}
			\toprule
			& \multicolumn{2}{c}{RHF-NK1} & \multicolumn{2}{c}{RHF-NK2} \\
			Parameter & w/o $\delta$ & with $\delta$ & w/o $\delta$ & with $\delta$ \\
			\midrule
			{$g_{\sigma}(0)$}&8.1814&-&7.7963&-\\
			$a_{\sigma}$&0.8964&-&0.9981&-\\
			$b_{\sigma}$&12.1538&-&15.1500&-\\
			$c_{\sigma}$&10.8202&-&15.1197&-\\
			$d_{\sigma}$&0.1755&-&0.1485&-\\
			$m_{\sigma}$ [MeV]&541.4493&-&535.0394&-\\
			{$g_{\omega}(0)$}&10.0446&9.0387&9.4614&8.5418\\
			$a_{\omega}$&0.9386&-&0.9991&-\\
			$b_{\omega}$&3.2379&-&2.9831&-\\
			$c_{\omega}$&3.0045&-&2.9798&-\\
			$d_{\omega}$&0.3331&-&0.3345&-\\
			{$g_{\rho}(0)$}&4.2595&7.4242&2.1604&6.9236\\
			$a_{\rho}$&0.9520&0.4833&0.6569&0.4103\\
			$f_{\rho}(0)/2M$ [fm]&1.1059&0.4467&0.9097&0.4831\\
			$a_{\rho f}$&0.2275&-&0.3831&-\\
			$f_{\pi}(0)$&1.3248&0.0028&1.4321&0.5851\\
			$a_{\pi}$&1.4845&-&0.5059&-\\
			{$g_{\delta}(0)$}&0.0000&5.4281&0.0000&5.7828\\
			$a_{\delta}$&--&0.0000&--&0.0000\\
			$\rho_{0}$ [fm$^{-3}$]&0.1550&-&0.1645&-\\
			
			\bottomrule

		\end{tabular}	
	\end{table}
	
	\par
	The final low-$L$ parametrizations have $a_\delta=0$, corresponding to a constant $\delta$ coupling. We also tested explicitly density-dependent $\delta$ couplings. Under the same saturation-property constraints, such parametrizations reduce $L$ less efficiently, and the optimization drives $a_\delta$ toward zero. The optimized low-$L$ solution is therefore characterized by a slowly varying isovector sector, with a constant $\delta$ coupling and a weakened density dependence of the $\rho$ coupling.
	\par

	The corresponding saturation properties are summarized in Table~\ref{tab:sat_property_delta}. The binding energy per nucleon $E/A$, saturation density $\rho_0$, symmetry energy $E_{\mathrm{sym}}$, and incompressibility coefficient $K$ remain nearly unchanged. In contrast, $L$ decreases from $73.72$ to $31.94~\mathrm{MeV}$ for RHF-NK1 and from $72.96$ to $31.99~\mathrm{MeV}$ for RHF-NK2. The changes in $Q_{\mathrm{sat}}$ and $K_{\mathrm{sym}}$ also indicate a softer high-density behavior, consistent with the trend favored by Bayesian inference~\cite{dong2025equation}. These results show that the $\delta$ meson opens a low-$L$ region of the DDRHF parameter space while preserving the constrained bulk properties.
	\par
	
	\begin{table}[!htbp]
		
		\centering
		
		\caption{Saturation properties of nuclear matter for RHF-NK1 and RHF-NK2 before and after the inclusion of the $\delta$-meson channel.}
		
		\label{tab:sat_property_delta}
		
		\footnotesize
		
		\setlength{\tabcolsep}{4pt}
		
		\renewcommand{\arraystretch}{1.05}
		
		\begin{tabular}{lcccc}
			
			\toprule
			
			& \multicolumn{2}{c}{ RHF-NK1} & \multicolumn{2}{c}{RHF-NK2} \\
			
			Property & w/o $\delta$ & with $\delta$ & w/o $\delta$ & with $\delta$ \\
			
			\midrule
			$\rho_0$ [fm$^{-3}$]        & 0.155   & 0.155   & 0.165   & 0.165   \\
			
			$E/A$ [MeV]                  & -15.823 & -15.830 & -15.877 & -15.879 \\
			
			$K$ [MeV]                   & 267.568 & 267.543 & 250.643 & 250.649 \\
			
			$Q_{\mathrm{sat}}$ [MeV]    & 324.377 & -407.518& 36.159  & -349.770\\
			
			$M^{*}/M$                   & 0.638   & 0.639   & 0.656   & 0.645   \\
			
			$E_{\mathrm{sym}}$ [MeV]    & 32.198  & 32.185  & 30.788  & 30.768  \\
			
			$L$ [MeV]                   & 73.717  & 31.935  & 72.960  & 31.990  \\
			
			$K_{\mathrm{sym}}$ [MeV]    & 110.090 & -49.675 & 68.545  & -59.242 \\
			
			\bottomrule
			
		\end{tabular}
		
	\end{table}
	
	\par
	% Insert Table II here:
	% Changes in the meson--nucleon coupling parameters.

	\par
	We next examine the microscopic origin of the reduced slope parameter through channel decompositions of the symmetry energy and the density-dependent slope function.
	\par
	\begin{figure}[t]
		\centering
		\includegraphics[width=0.6\columnwidth]{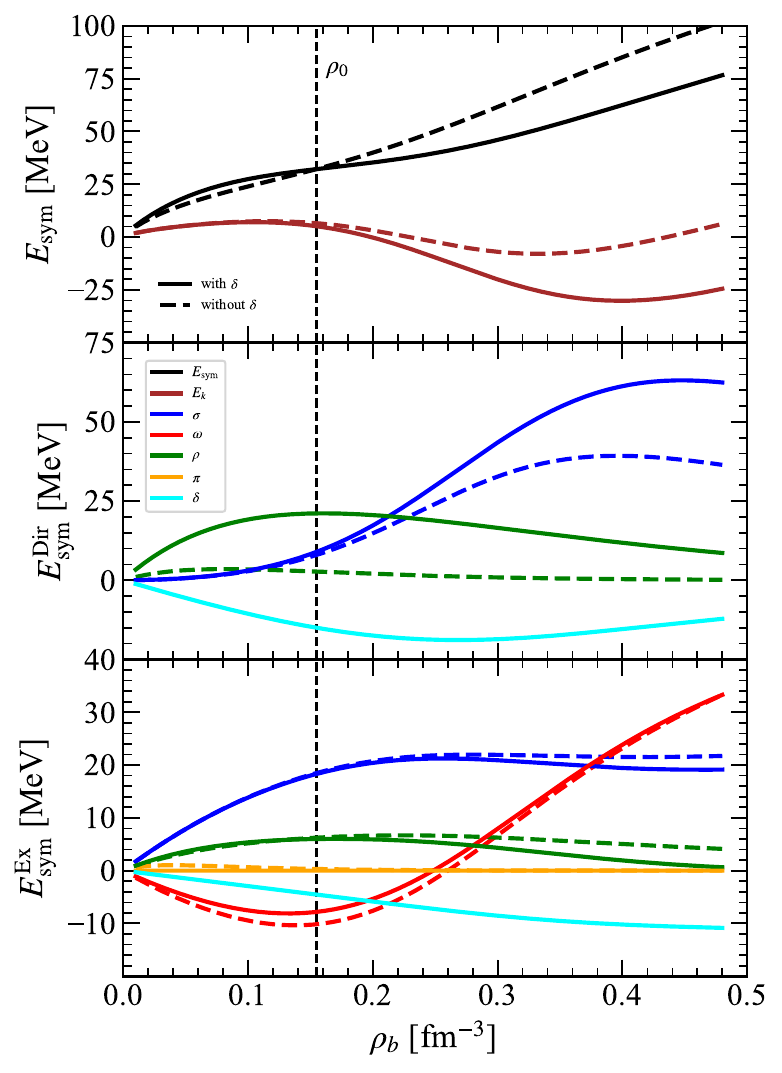}
		\caption{
			Contributions of the different meson-exchange
			channels to the symmetry energy for RHF-NK1.
			Solid and dashed lines denote the results with and
			without the $\delta$-meson channel, respectively.
		}
		\label{fig:symmetry_energy_decomposition}
	\end{figure}
	\par
	Fig.~\ref{fig:symmetry_energy_decomposition} shows the channel decomposition of the symmetry energy for RHF-NK1. The kinetic, Hartree, and Fock contributions are displayed separately. The $\sigma$ and $\omega$ mesons give sizable contributions despite their isoscalar character, consistent with Ref.~\cite{zhao2015kin,miyatsu2020decomposition}.
	\par
	Below and around saturation density, the total symmetry energy is slightly enhanced after the $\delta$ meson is included, mainly because of the larger $\rho$ contribution. The negative $\delta$ contribution moderates this enhancement. Above saturation density, the total symmetry energy is reduced by the combined decrease of the kinetic term, the $\sigma$-Fock contribution, and the negative $\delta$ contribution, while the enhanced $\rho$ contribution partially compensates for the reduction. The optimized $\delta$ channel therefore bends $E_{\mathrm{sym}}(\rho_b)$ downward around saturation density and reduces the corresponding slope parameter.
	\par
	\begin{figure}[t]
		\centering
		\includegraphics[width=0.6\columnwidth]{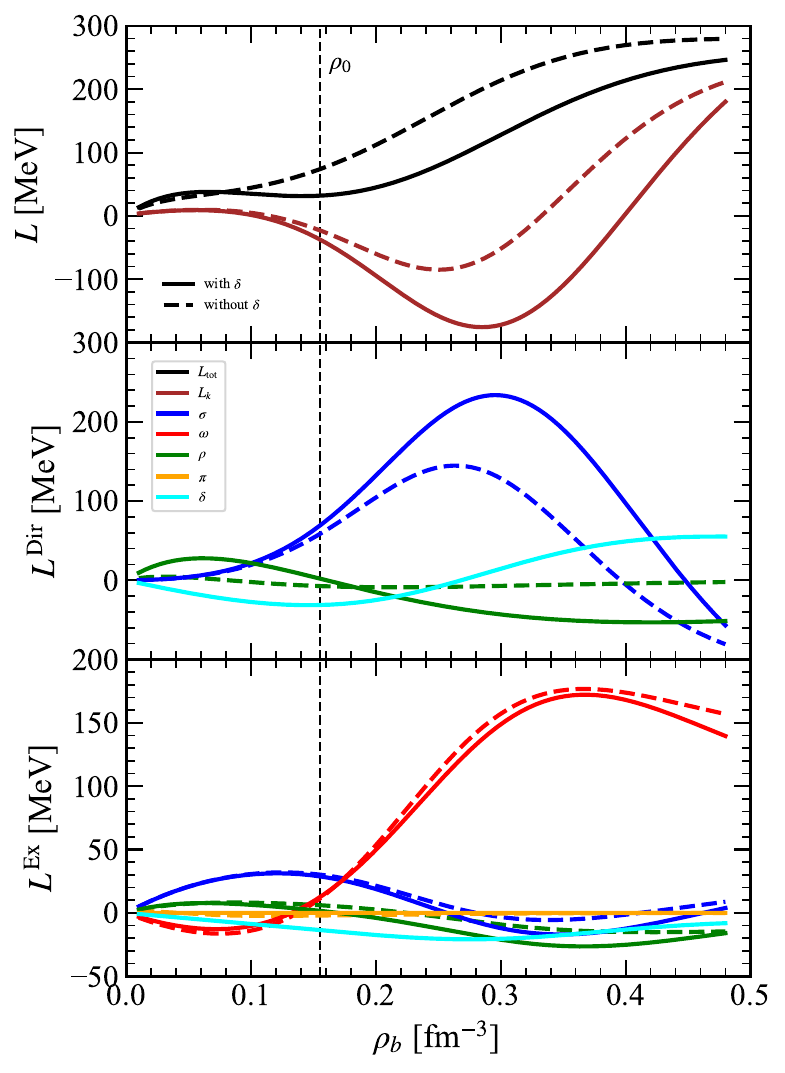}
		\caption{
			Contributions of the different interaction channels
			to the density-dependent slope function $L(\rho_b)$ defined in Eq.~\eqref{eq:Lrho} for RHF-NK1.
		}
		\label{fig:slope_decomposition}
	\end{figure}
	\par
	To quantify the change in the density dependence of the symmetry energy, we decompose the density-dependent slope function defined in Eq.~\eqref{eq:Lrho} into contributions from the individual interaction channels for RHF-NK1. The corresponding results are shown in Fig.~\ref{fig:slope_decomposition}. Around and below saturation density, the total slope is reduced after the inclusion of the $\delta$ meson, indicating a softer density dependence. This reduction is primarily driven by the decrease in the kinetic contribution. The increased contribution from the $\sigma$-meson channel partly compensates for the reduction in the kinetic term, while the changes in the $\rho$-meson contribution and the additional $\delta$-meson contribution provide further adjustments. The reduction of the saturation-density value of $L$ therefore results from a coordinated rearrangement of contributions from different interaction channels. The readjustment of the meson--nucleon couplings not only reduces $L$, but also preserves the constrained bulk properties of nuclear matter, allowing the role of the optimized $\delta$ channel to be identified more clearly.

	\paragraph{Comparison with previous RMF and RHF studies}
	The present mechanism can be contrasted with earlier RMF studies of the isovector-scalar channel. In conventional RMF models, the direct $\delta$ contribution to the symmetry energy is negative, but keeping $E_{\mathrm{sym}}(\rho_0)$ fixed usually requires a stronger $\rho$ coupling. This compensation often increases the high-density neutron-matter energy, proton fraction, and stiffness of $\beta$-stable matter~\cite{kubis1997nuclear,kubis1998neutron,liu2002asymmetric,gaitanos2004lorentz,wang2014neutron}. DD-ME$\delta$ and more recent Bayesian, covariance, and scalar-mixing analyses further show that the $\delta$ meson can enlarge the accessible symmetry-energy parameter space and affect neutron-star radii, tidal deformabilities, and proton fractions~\cite{rocamaza2011relativistic,thakur2022effects,santos2025impact,li2022effects,kumar2023implications}.
	
	The DDRHF result obtained here differs from this conventional RMF trend. After the nearly density-independent $\delta$ channel is introduced and the couplings are refitted, the high-density symmetry energy and neutron-star radii are reduced. This softening should not be interpreted as the direct Hartree effect of the $\delta$ meson alone. Instead, it reflects a correlated rearrangement of the $\delta$, $\rho$, $\pi$, and isoscalar Fock contributions from $\sigma$ and $\omega$ mesons. This interpretation is consistent with previous DDRHF and Lorentz-covariant RHF decompositions, which showed that isoscalar exchange terms and Fock contributions play important roles in $E_{\mathrm{sym}}$ and its density dependence~\cite{sun2008neutron,miyatsu2020decomposition}.
	
\begin{figure}[t]
		\centering
		\includegraphics[width=0.6\columnwidth]{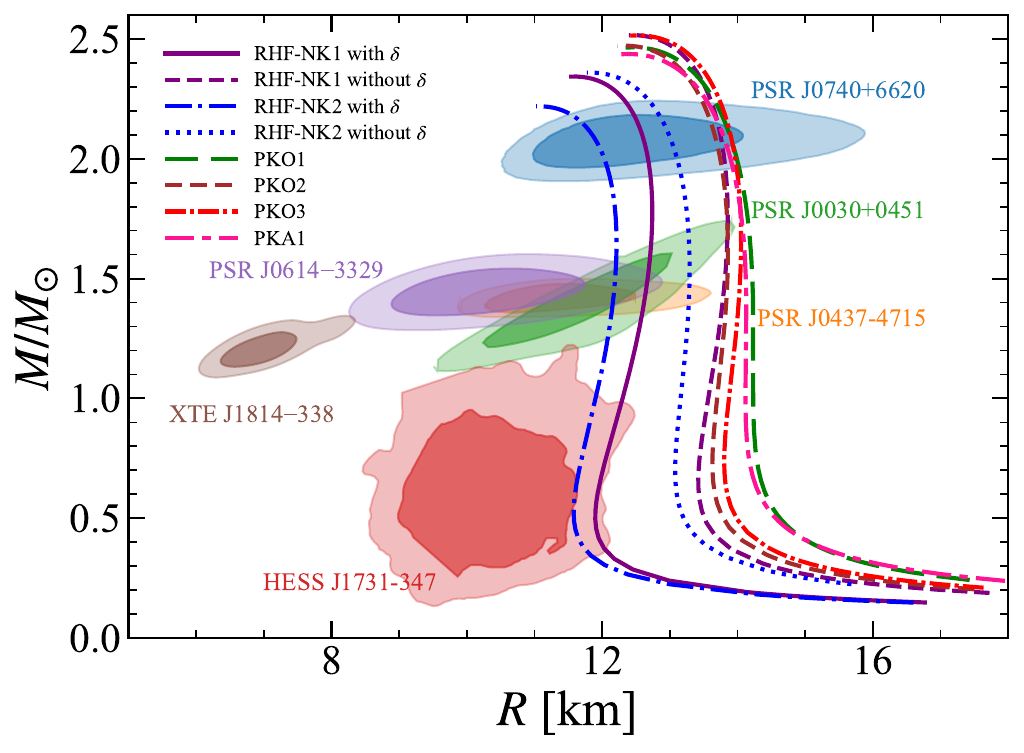}
		\caption{
			Neutron-star mass--radius relations obtained with RHF-NK1
			and RHF-NK2 before and after the inclusion of
			the $\delta$-meson channel. The results of the existing
			PKO1, PKO2, PKO3, and PKA1 DDRHF parametrizations are
			shown for comparison, together with representative
			astrophysical mass--radius constraints.
		}
		\label{fig:mr_with_delta}
	\end{figure}

	We finally examine the impact of the $\delta$ meson on neutron-star structure in the DDRHF model. Fig.~\ref{fig:mr_with_delta} shows the mass--radius relations for RHF-NK1 and RHF-NK2 before and after the inclusion of the $\delta$ meson. The curves shift to smaller radii after the low-$L$ refit, while the maximum masses are also reduced. For RHF-NK1 and RHF-NK2, $R_{1.4}$ changes from $13.78$ and $13.28~\mathrm{km}$ to $12.62$ and $12.16~\mathrm{km}$, respectively. The optimized $\delta$ channel therefore provides an effective mechanism for reducing the radii predicted by conventional DDRHF parametrizations.
	\par
	% Insert Fig. 4 here:
	% Neutron-star mass--radius relations before and after
	% the inclusion of the delta-meson channel.
	
	\section{Summary and conclusions}
	In this work, we incorporated the isovector-scalar $\delta$ meson into the DDRHF framework and examined its impact on the symmetry energy and neutron-star radii. As a no-$\delta$ baseline, we generated $1006$ DDRHF parametrizations satisfying broad saturation-property constraints. Their symmetry-energy slope parameters lie in the range $L\simeq65$--$110~\mathrm{MeV}$, with no accepted set reaching the low-$L$ region. Two representative parametrizations, RHF-NK1 and RHF-NK2, were then used to examine the effect of the $\delta$ channel. After the $\delta$ meson is introduced, the optimization selects a density-independent $\delta$ coupling constant in the final low-$L$ parametrizations. The slope parameter is reduced from approximately $73$ to $32~\mathrm{MeV}$, while the binding energy per nucleon, saturation density, symmetry energy, and incompressibility coefficient remain nearly unchanged.
	\par
	The channel decomposition shows that this softening is not caused by the direct $\delta$ contribution alone. It results from a coordinated readjustment of the $\delta$, $\rho$, $\pi$, and isoscalar Fock contributions, which bends the symmetry energy downward around and above saturation density. The predicted radii of $1.4M_\odot$ neutron stars, $R_{1.4}$, decrease from 13.78 and 13.28 km to 12.62 and 12.16 km for RHF-NK1 and RHF-NK2, respectively. They are broadly compatible with recent NICER and multimessenger mass--radius constraints. These results indicate that the $\delta$ meson provides an efficient additional degree of freedom in the DDRHF isovector sector. The preference for an almost constant $\delta$ coupling emerges from the numerical search for low-$L$ solutions rather than being assumed at the outset.
	
	{An extension of the present low-\(L\) DDRHF parametrizations to finite nuclei is
	an important next step. Recent studies have shown that the isovector spin-orbit
	potential may play an important role in neutron-rich nuclei and in the
	CREX-PREX dilemma~\cite{Kunjipurayil2025PRC}. The \(\delta\) channel and the
	associated exchange terms may also affect neutron skins, spin-orbit splittings,
	and other isovector observables in finite nuclei. Such calculations are more
	involved than the present nuclear-matter study and are currently in progress.}

	\section*{Acknowledgments}
	This work was supported by the National Natural Science Foundation of China under Grant No.~12475149 and by the Guangdong Basic and Applied Basic Research Foundation under Grant No.~2024A1515010911.
	
	
	

	%\bibliographystyle{elsarticle-harv}
	
	\bibliographystyle{elsarticle-num}

    \bibliography{reference}
	
	
	
	%% For example:
	
	%% \bibitem[Aladro et al.(2015)]{Aladro15} Aladro, R., Martín, S., Riquelme, D., et al. 2015, \aas, 579, A101

	%%\end{thebibliography}
	
\end{document}